\pdfoutput=1 
\documentclass[aps,prd,superscriptaddress,tightenlines,nofootinbib,preprint]{revtex4}

\usepackage{graphicx}
\usepackage{rotating}
\usepackage{epsfig}
\usepackage{bm}
\usepackage{latexsym,amssymb,amsmath,amsfonts,amssymb,txfonts,pxfonts,wasysym,float}
\usepackage{multirow}
\usepackage{enumitem}
\usepackage{color}
\newcommand{\beq}[1]{\begin{equation}\label{#1}}
\newcommand{\eeq}{\end{equation}}
\newcommand{\bea}[1]{\begin{eqnarray} \label{#1}}
\newcommand{\eea}{\end{eqnarray}}
\newcommand{\ba}{\begin{array}}
\newcommand{\ea}{\end{array}}

\def\be{\begin{equation}}
\def\ee{\end{equation}}
\def\gs{\mathrel{
   \rlap{\raise 0.511ex \hbox{$>$}}{\lower 0.511ex \hbox{$\sim$}}}}
\def\ls{\mathrel{
   \rlap{\raise 0.511ex \hbox{$<$}}{\lower 0.511ex \hbox{$\sim$}}}}

\newcommand{\postscript}[2]{\setlength{\epsfxsize}{#2\hsize}
   \centerline{\epsfbox{#1}}}

\newcommand{\comment}[1]{}

\usepackage[usenames,dvipsnames]{xcolor}
\definecolor{orange}{cmyk}{0,0.5,1,0}
\definecolor{rossoCP3}{cmyk}{0,.88,.77,.40}
\definecolor{graa}{rgb}{0.8,0.8,0.8}
\definecolor{blaa}{rgb}{0.2,0.2,0.6}

\begin{document}

\pagenumbering{gobble}
\clearpage
\thispagestyle{empty}

\title{POEMMA (Probe of Extreme Multi-Messenger Astrophysics) design}
\author{A V. Olinto}
\affiliation{The University of Chicago, Chicago, IL, USA,}
\author{J. H. Adams}
\affiliation{University of Alabama, Huntsville, AL, USA}
\author{R. Aloisio}
\affiliation{Gran Sasso Science Institute, L'Aquila, Italy}
\author{L. A. Anchordoqui}
\affiliation{City University of New York, Lehman College, NY, USA}
\author{D. R. Bergman}
\affiliation{University of Utah, Salt Lake City, Utah, USA}
\author{M. E. Bertaina}
\affiliation{Universit\'a di Torino, Torino, Italy}
\author{P. Bertone}
\affiliation{NASA Marshall Space Flight Center, Huntsville, AL, USA}
\author{F. Bisconti}
\affiliation{Universit\'a di Torino, Torino, Italy}
\author{M. Bustamante}
\affiliation{Niels Bohr Institute, University of Copenhagen, Denmark}
\author{M. Casolino}
\affiliation{RIKEN, Wako, Japan}
\author{M. J. Christl}
\affiliation{NASA Marshall Space Flight Center, Huntsville, AL, USA}
\author{A. L. Cummings}
\affiliation{Gran Sasso Science Institute, L'Aquila, Italy}
\author{I. De Mitri}
\affiliation{Gran Sasso Science Institute, L'Aquila, Italy}
\author{R. Diesing}
\affiliation{The University of Chicago, Chicago, IL, USA}
\author{J. Eser}
\affiliation{Colorado School of Mines, Golden, CO, USA}
\author{F. Fenu}
\affiliation{Universit\'a di Torino, Torino, Italy}
\author{C. Guepin}
\affiliation{Sorbonne Universit\'es, Institut dAstrophysique de Paris, Paris, France}
\author{E. A. Hays}
\affiliation{NASA Goddard Space Flight Center, Greenbelt, MD, USA}
\author{E. G. Judd}
\affiliation{Space Sciences Laboratory, University of California, Berkeley, CA, USA}
\author{J. F. Krizmanic}
\affiliation{NASA Goddard Space Flight Center, Greenbelt, MD, USA}
\author{E. Kuznetsov}
\affiliation{University of Alabama, Huntsville, AL, USA}
\author{A. Liberatore}
\affiliation{Universit\'a di Torino, Torino, Italy}
\author{S. Mackovjak}
\affiliation{Institute of Experimental Physics, Slovak Academy of Sciences,
Kosice, Slovakia}
\author{J. McEnery}
\affiliation{NASA Goddard Space Flight Center, Greenbelt, MD, USA}
\author{J. W. Mitchell}
\affiliation{NASA Goddard Space Flight Center, Greenbelt, MD, USA}
\author{A. Neronov}
\affiliation{University of Geneva, Geneva, Switzerland}
\author{F. Oikonomou}
\affiliation{European Southern Observatory, Garching bei M\"unchen, Germany}
\author{A. N. Otte}
\affiliation{Georgia Institute of Technology, Atlanta, GA, USA}
\author{E. Parizot}
\affiliation{APC, U. Paris Diderot, CNRS, CEA, Paris, France}
\author{T. Paul}
\affiliation{City University of New York, Lehman College, NY, USA}
\author{J. S. Perkins}
\affiliation{NASA Goddard Space Flight Center, Greenbelt, MD, USA}
\author{G. Pr\'ev\^ot}
\affiliation{APC, U. Paris Diderot, CNRS, CEA, Paris, France}
\author{P. Reardon}
\affiliation{University of Alabama, Huntsville, AL, USA}
\author{M. H. Reno}
\affiliation{University of Iowa, Iowa City, IA, USA}
\author{M. Ricci}
\affiliation{Istituto Nazionale di Fisica Nucleare - Laboratori Nazionali di Frascati, Frascati, Italy}
\author{F. Sarazin}
\affiliation{Colorado School of Mines, Golden, CO, USA}
\author{K. Shinozaki}
\affiliation{Universit\'a di Torino, Torino, Italy}
\author{J. F. Soriano}
\affiliation{City University of New York, Lehman College, NY, USA}
\author{F. Stecker}
\affiliation{NASA Goddard Space Flight Center, Greenbelt, MD, USA}
\author{Y. Takizawa}
\affiliation{RIKEN, Wako, Japan}
\author{R. Ulrich}
\affiliation{Karlsruhe Institute of Technology, Karlsruhe, Germany}
\author{M. Unger}
\affiliation{Karlsruhe Institute of Technology, Karlsruhe, Germany}
\author{T. M. Venters}
\affiliation{NASA Goddard Space Flight Center, Greenbelt, MD, USA}
\author{L. Wiencke}
\affiliation{Colorado School of Mines, Golden, CO, USA}
\author{R. M. Young}
\affiliation{NASA Marshall Space Flight Center, Huntsville, AL, USA}

\begin{abstract}
The Probe Of Extreme Multi-Messenger Astrophysics (POEMMA) is a NASA Astrophysics probe-class mission designed to observe ultra-high energy cosmic rays (UHECRs) and cosmic neutrinos from space.  \\
\\
Lead Author: Angela V. Olinto, contact: aolinto@uchicago.edu\\
\\
\noindent {\bf Astro2020 APC white paper: Medium-class Space Particle Astrophysics Project \\
Science White papers on Compact Objects and Energetic Phenomena; Galaxies; Cosmology.} 
{\it Relevant white papers submitted include (NAS response ID): 107, 147, 206, 239, 253, 263, 265, 272, 275, 277, 377, 439} 
\end{abstract}
\maketitle

\clearpage
\pagenumbering{arabic}

\section{POEMMA Overview}

The Probe Of Extreme Multi-Messenger Astrophysics (POEMMA) is a NASA Astrophysics probe-class mission designed to observe ultra-high energy cosmic rays (UHECRs) and cosmic neutrinos from space. POEMMA will monitor colossal volumes of the Earth's atmosphere to detect extensive air showers (EASs) produced by extremely energetic cosmic messengers,  cosmic neutrinos above  20 PeV and UHECRs above 20 EeV, over the entire sky. The POEMMA conceptual study report prepared for NASA is here~\cite{NASApoemma}.

\begin{figure}[tbp]
    \postscript{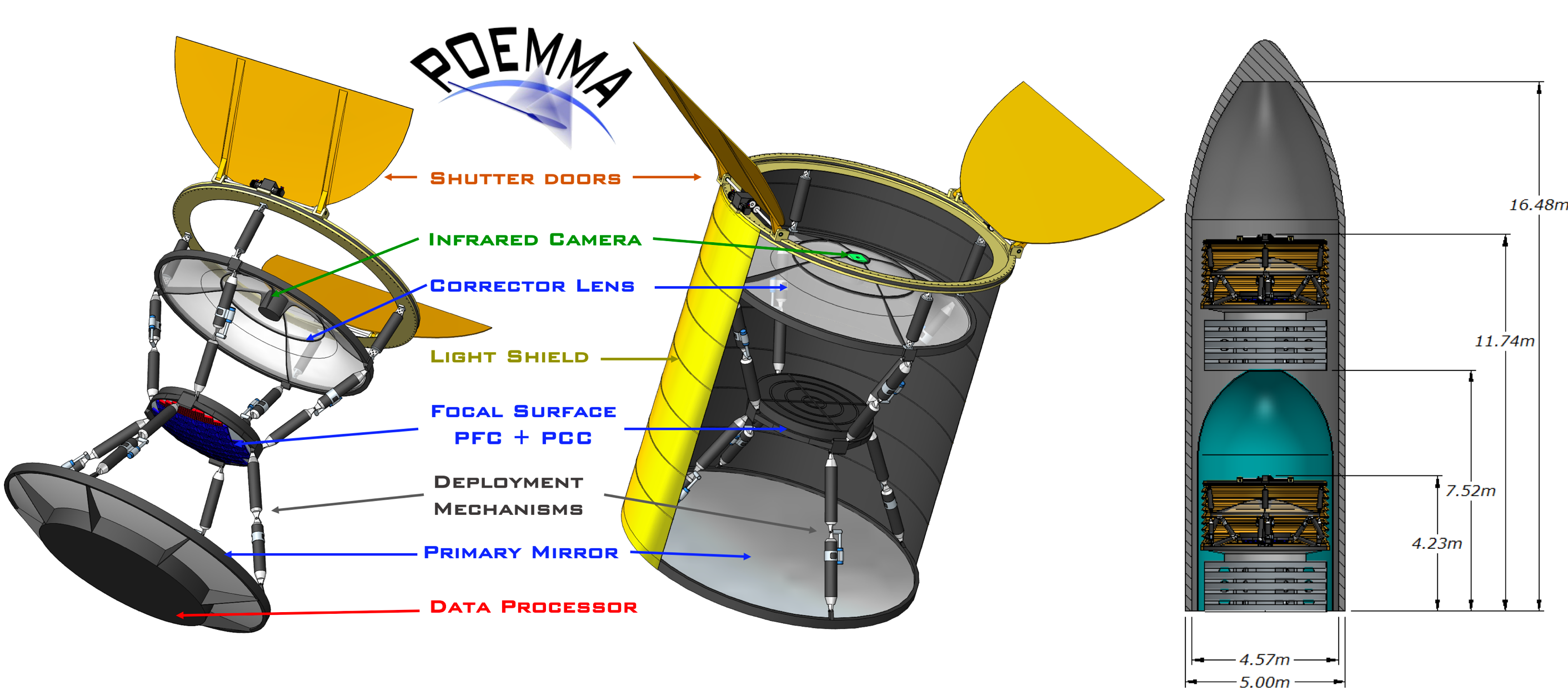}{1}
    \caption{Left: Concept of the POEMMA photometer with major components identified. Right: Both POEMMA photometers accommodated on Atlas V for launch. \label{fig1}}
\end{figure}

POEMMA is comprised of two identical observatories flying in loose formation to detect EASs in mono and stereo modes. Each observatory is composed of a 4-meter photometer designed with Schmidt wide (45$^{\circ}$) field-of-view (FoV) optics (see Fig. \ref{fig1}) and a spacecraft bus. The photometer focal surface has a hybrid design for two complementary capabilities: a fast (1 $\mu$s) ultraviolet camera to observe fluorescence signals and an ultrafast (20 ns) optical camera to detect Cherenkov signals. EASs from UHECRs and cosmic neutrinos are observed from an orbit altitude of 525 km and a range of attitudes in the dark sky. POEMMA will point from close to the nadir, to optimize stereo fluorescence observations, to about 47$^\circ$ from the nadir to monitor the Earth's limb (located at 67.5$^\circ$) for Cherenkov emission from cosmic tau-neutrino induced EASs (below the limb) and UHECRs ($\sim 2^\circ$ above the limb). 

POEMMA will provide a significant increase in the statistics of observed UHECRs at the highest energies over the entire sky and will pioneer a target of opportunity (ToO) follow-up program for cosmic neutrinos from extremely energetic transient astrophysical events. 

\medskip
\noindent {\bf POEMMA} is designed to:

$\bullet$ {\bf Discover the nature and origin of the highest-energy particles in the universe.} Where do UHECRs come from?  What are these extreme cosmic accelerators and how do they accelerate to such high energies? What is the UHECR composition at the highest energies? What are the magnetic fields in the galactic and extragalactic media? How do UHECRs interact in the source, in galactic and extragalactic space, and in the atmosphere of the Earth? 

$\bullet$ {\bf Discover neutrino emission from extreme astrophysical transients above 20 PeV.} What is the high-energy neutrino emission of gravitational wave events? Do binaries with black holes, white dwarfs, and neutron stars produce high-energy neutrinos when they coalesce? Neutrino observations will elucidate the underlying dynamics of blazar flares, gravitational wave events, gamma-ray bursts, newborn pulsars, tidal disruption events, and other transient events as seen by neutrinos.

$\bullet$ {\bf Probe particle interactions at extreme energies.} POEMMA can test models with physics beyond the Standard Model (BSM) through cosmic neutrino observations from tens of PeVs to tens of ZeVs;

$\bullet$ {\bf Observe Transient Luminous Events} and study the dynamics of the Earth's atmosphere, including extreme storms;

$\bullet$ {\bf Observe Meteors}, thereby contributing to the understanding of the dynamics of meteors in the Solar System;

$\bullet$ {\bf Search for Exotic particles} such as nuclearites.

\medskip

POEMMA will provide new {\bf Multi-Messenger Windows} onto the most energetic environments and events in the universe, enabling the study of new astrophysics and  particle physics at these otherwise inaccessible energies.

\vfill 
\eject

\begin{table}
\centering
\caption{POEMMA Specifications:}
\label{tab-1}  \
\begin{tabular}{lllllllll}
\hline
\hline
Photometer & Components &  &$\ \ $& Spacecraft  & \\ \hline
Optics &  Schmidt & 45$^\circ$ full FoV && Slew rate & 90$^\circ$  in 8 min \\
 & Primary Mirror & 4 m diam. && Pointing Res. & 0.1$^\circ$ \\
 & Corrector Lens & 3.3 m diam. && Pointing Know. & 0.01$^\circ$ \\
 & Focal Surface & 1.6 m diam. && Clock synch. & 10 ns \\  
 & Pixel Size & $3 \times 3$ mm$^2$  && Data Storage & 7 days \\ 
& Pixel FoV & 0.084$^\circ$ && Communication & S-band \\ 
PFC & MAPMT (1$\mu$s)& 126,720 pixels  && Wet Mass & 3,450 kg \\
PCC & SiPM (20 ns)& 15,360 pixels  && Power (w/cont)& 550 W \\ \hline
Photometer & (One)&  &&Mission  &(2 Observatories) \\ \hline
 & Mass & 1,550 kg  && Lifetime & 3 year  (5 year goal)\\
 & Power (w/cont) & 700 W   && Orbit & 525 km, 28.5$^\circ$ Inc \\
 & Data & $<$ 1 GB/day && Orbit Period & 95 min \\
& &  && Observatory Sep. & $\sim$25 - 1000 km \\\hline
\hline
\end{tabular}
 \
\center{Each Observatory = Photometer + Spacecraft; POEMMA Mission = 2 Observatories}

\end{table}

\section{POEMMA Extreme Multi-Messenger Science}

The main scientific goals of  POEMMA  are to discover the elusive sources of UHECRs, cosmic rays with energies above 10$^{18}$ eV ($\equiv$ 1 EeV), and to observe cosmic neutrinos from multi-messenger transients. POEMMA exploits the tremendous gains in both UHECR and cosmic neutrino exposures offered by space-based measurements, including {\it full-sky coverage} of the celestial sphere.  For cosmic rays with energies $E \agt 20~{\rm EeV}$, POEMMA will enable charged-particle astronomy by obtaining definitive measurements of the UHECR spectrum,  composition, and source location. For multi-messenger transients, POEMMA will follow-up targets of opportunity transients to detect the first cosmic neutrino emission with energies $E_{\nu} \agt$ 20 PeV. POEMMA also has sensitivity to neutrinos with energies above 20~EeV through fluorescence observations of neutrino induced EASs. 
Supplementary science capabilities of POEMMA include probes of physics beyond the Standard Model of particle physics, the measurement of $pp$ cross-section at 238 TeV center-of-mass energy, the study of atmospheric transient luminous events (TLEs), and the search for meteors and nuclearites (see \cite{NASApoemma,UHECRpoemma} for more details).

\begin{figure}[ht]
    \postscript{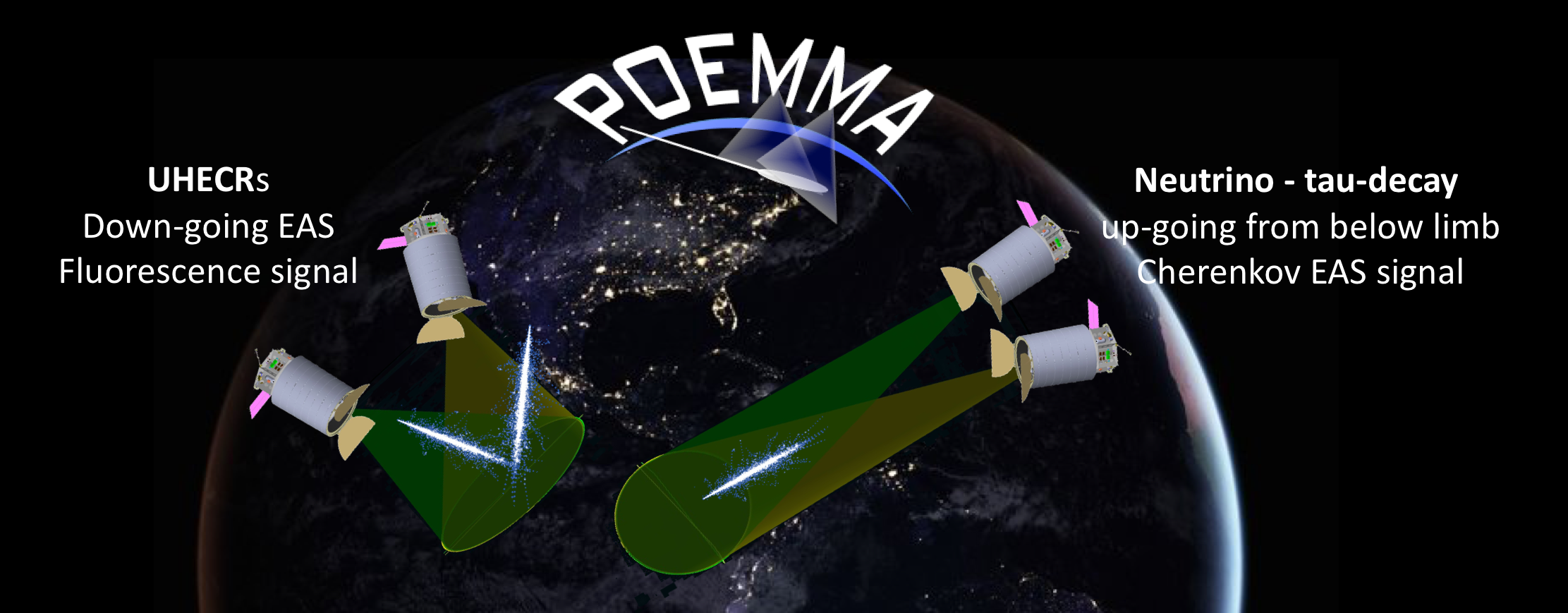}{0.9}
\caption{POEMMA observing modes. Left:  Stereo fluorescence mode around the nadir. Right: Cherenkov mode from above (UHECRs) and below (cosmic tau neutrinos) the limb of the Earth; 
\label{fig2}}
\end{figure}

These groundbreaking measurements are obtained by operating POEMMA's two observatories (described in Fig. \ref{fig1} and Table I) in different orientation modes: a quasi-nadir stereo fluorescence configuration for UHECR observations and a tilted, Earth-limb viewing Cherenkov configuration for ToO neutrino searches (see Fig. \ref{fig2}). In limb observing mode, POEMMA can simultaneously search for neutrinos and UHECRs with Cherenkov observations, while observing UHECRs with fluorescence, thanks to the POEMMA hybrid camera (or Focal Surface) design.  

In stereo mode, the two photometers with several square meters of collecting area view a common, immense volume of  $\sim 10^{4}$ gigatons of atmosphere. POEMMA's fluorescence observations yield one order of magnitude increase in yearly UHECR exposure compared to ground observatory arrays and two orders of magnitude compared to ground fluorescence telescopes. In the Cherenkov limb-viewing mode, POEMMA searches for optical Cherenkov signals of upward-moving EASs generated by $\tau$-lepton decays produced by $\nu_\tau$ interactions in the Earth. The terrestrial neutrino target monitored by POEMMA  reaches  $\sim 10^{10}$ gigatons. 
In the Cherenkov mode, an even more extensive  volume is monitored for UHECR fluorescence observations. Thus, {\bf POEMMA uses  the Earth and its atmosphere as a gargantuan high-energy physics detector and astrophysics observatory.}

\bigskip

\noindent{\bf IIa. UHECR Science}

\begin{figure}[ht]
    \postscript{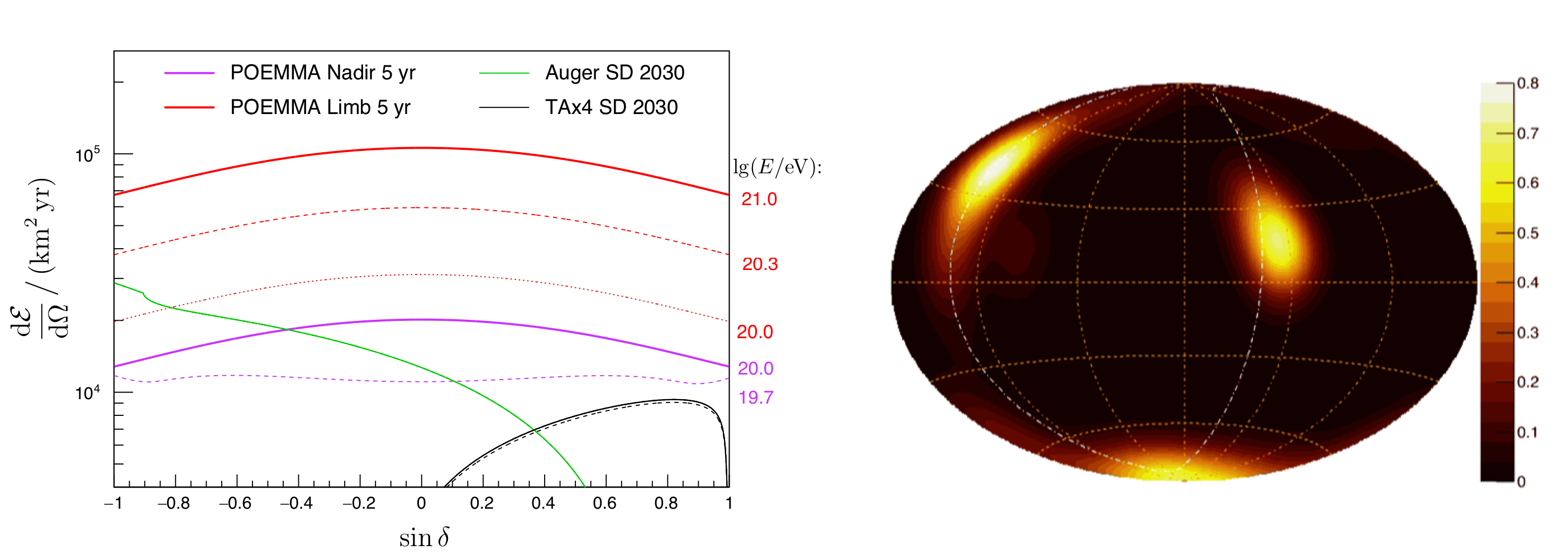}{0.99}
  \caption{Left: Differential exposure vs declination for POEMMA 5-yr in nadir  (purple) at 10$^{19.7}$ eV (dotted) and 10$^{20}$ eV (solid); and for limb (red) at 10$^{20}$ eV (dotted), 10$^{20.3}$ eV (dashed), and 10$^{21}$ eV (solid).  Exposures for Auger (green) and TAx4  (black) surface detectors (SD)  until 2030.
  Right: Skymap of nearby starburst galaxies from~\cite{Ackermann:2012vca} with POEMMA exposure with $12.9^{\circ}$ smoothing as in~\cite{Aab:2018chp}. The color scale indicates the probability density of the source sky map, as a function of position on the sky. (Supergalactic Plane in white dashed line.) From~\cite{UHECRpoemma}. \label{figUHECR1}}
\end{figure}

A detailed study of the powerful UHECR science of POEMMA is described in~\cite{UHECRpoemma}.  The two leading UHECR observatories currently in operation are the Pierre Auger
Observatory~\cite{Abraham:2010zz,Abraham:2009pm} in the southern
hemisphere, with $\sim 6.5 \times 10^4~{\rm km^2 \ sr \ yr}$ exposure over 13
years~\cite{Aab:2017njo}, and the Telescope Array (TA)~\cite{AbuZayyad:2012kk,Tokuno:2012mi} in the northern hemisphere, with $\sim 10^4~{\rm km}^2 \, {\rm sr} \, {\rm yr}$ exposure in 10 years (TA is currently upgrading by a factor of 4, named TAx4). POEMMA can reach an order of magnitude increase in exposure compared to these ground arrays (from $\sim10^5$ to $ 10^6 ~{\rm km}^2 \, {\rm sr} \, {\rm yr}$ exposure in a 5 year mission) and two orders of magnitude when compared with their ground fluorescence capabilities ($\sim$ 10\% duty cycle for fluorescence on the ground). POEMMA observes the full sky with the same instrument. Fig.~\ref{figUHECR1} left shows POEMMA's differential exposure as a function of declination compared to Auger and TAx4 extrapolated to 2030~\cite{UHECRpoemma}. 

\begin{figure}[ht]
    \postscript{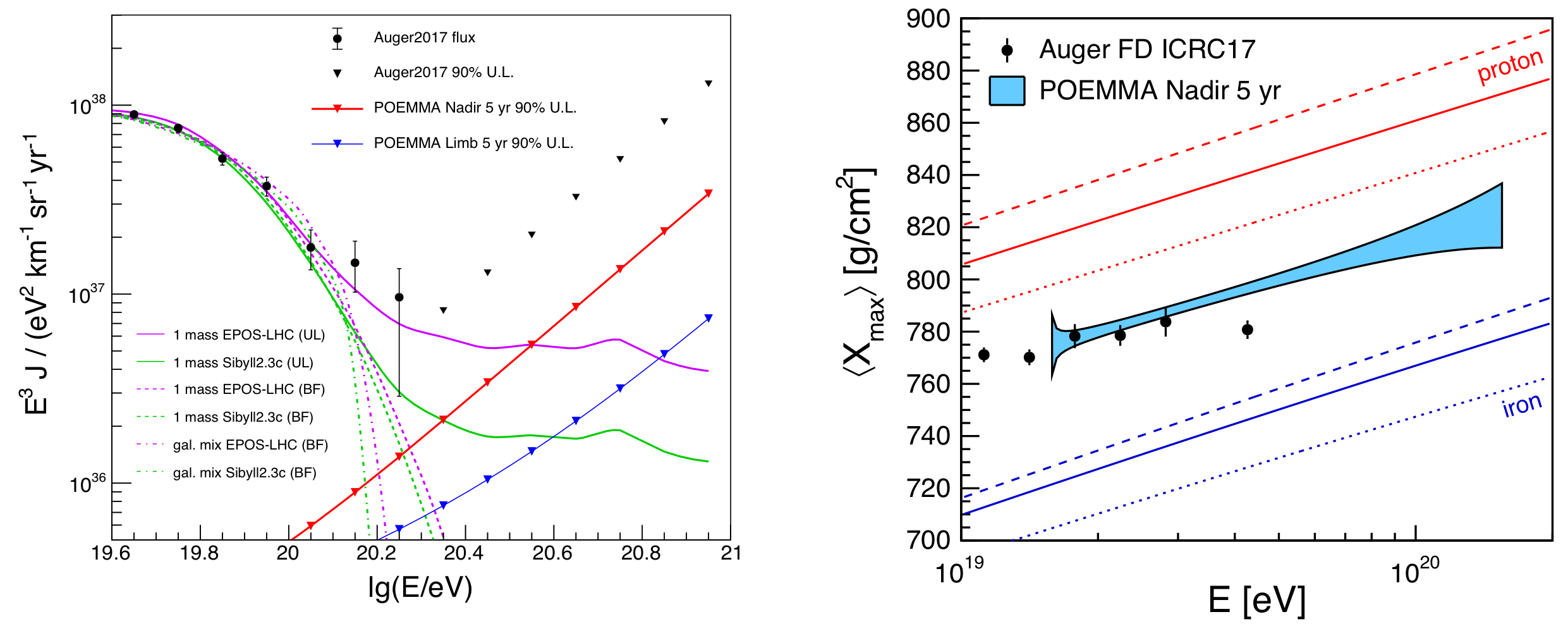}{0.9}
  \caption{Left: UHECR flux model predictions from~\cite{Muzio:2019leu} and Auger data~\cite{Aab:2017njo}. 90\% confidence upper limits are downward triangles: Auger 2017 (black), POEMMA 5 year stereo (red) and
  limb mode (blue).
     Right:  POEMMA capabilities for composition studies through measurements of the atmospheric column depth at which the EAS longitudinal development reaches maximum, $\langle X_{\rm max} \rangle$. 
The width of the blue band illustrates the expected statistical uncertainties. From~\cite{UHECRpoemma}.
  \label{figUHECR2}}
\end{figure}

The powerful increase in exposure provided by POEMMA will determine the sources of UHECRs  through the combined sensitive observations of the sky distribution, the spectrum, and the composition at the highest energies. Fig.~\ref{figUHECR1} right shows an example of a POEMMA sky distribution of events if UHECR sources are located in nearby starburst galaxies.
POEMMA will  explore the differences in source models for the UHECR predicted spectrum above the current reach in energy, as shown in Fig.~\ref{figUHECR2} left.  In addition, POEMMA will measure the UHECR composition at 100s of EeV where models differ in predictions (Fig.~\ref{figUHECR2} right) further illuminating the origin of UHECRs.

The nature of the astrophysical sources of UHECRs and their acceleration mechanism(s) remains a mystery as reviewed here~\cite{Kotera:2011cp,Anchordoqui:2018qom,Batista:2019}.  Proposed sources span a large range of astrophysical objects including extremely fast-spinning young pulsars~\cite{Blasi:2000xm,Fang:2012rx,Fang:2013cba}, active galactic nuclei (AGN)~\cite{Biermann:1987ep,Rachen:1992pg,Eichmann:2017iyr},
starburst galaxies (SBGs)~\cite{Anchordoqui:1999cu,Anchordoqui:2018vji}, and
gamma-ray bursts (GRBs)~\cite{Waxman:1995vg,Vietri:1995hs}. POEMMA data can determine which of the proposed model, if any, is the answer to this long-standing mystery.

\bigskip

\noindent{\bf IIb. Cosmic Neutrino Science}

 In addition to observing the components of UHECRs, the POEMMA design adds the novel capability of searching for neutrinos above 20 PeV (1 PeV $\equiv$ 10$^{15}$ eV)  from ToO events. Detailed studies of the POEMMA neutrino observing capabilities can be found in~\cite{Guepin2018,Reno2019a,Venters:2019xwi}. (We use neutrinos to denote both neutrinos and anti-neutrinos, which interact similarly at these energies.)

POEMMA is unique in the ability to follow-up transient events on time scales of order one orbit (95 min) over the entire dark sky and on time scales of months over the full sky. Since no prompt neutrinos have been observed at these energy scales, POEMMA will {\it discover} which transient events produce very-high energy neutrinos and at what times after the event. 

Very high-energy cosmic neutrinos are emitted in a number of models of astrophysical transient events~\cite{Meszaros:2017fcs,Ackermann:2019ows}. Astrophysical sources generally produce electron and muon neutrinos, which after astronomical propagation distances,  arrive on Earth with approximately equal numbers of the three flavors: electron, muon, and tau neutrinos. POEMMA detects primarily tau-neutrinos through the tau-decay generated EASs.  POEMMA will observe in general one third of the generated neutrino flux via the $\nu_\tau$ flux.

\begin{figure}[ht]
    \postscript{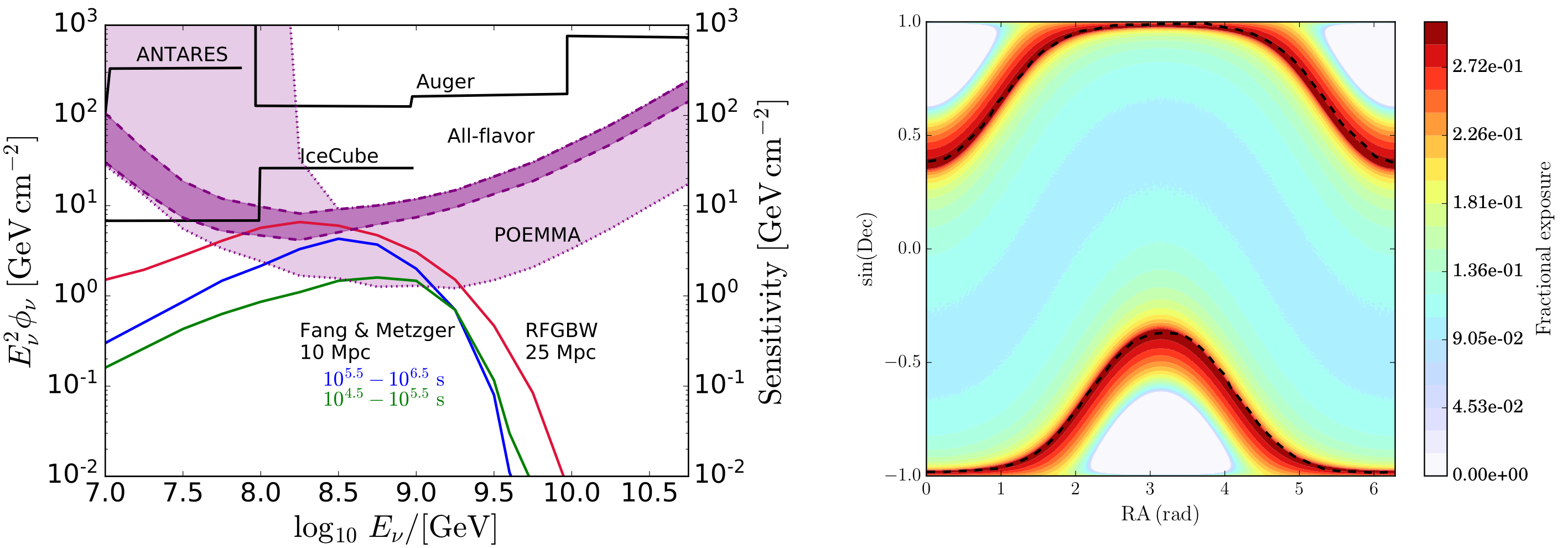}{1}
  \caption{Left: POEMMA ToO sensitivities to long bursts shown in purple. Dark
purple bands correspond to source locations between the dashed curves in the sky coverage figure on the right. Also shown IceCube, ANTARES, and Auger sensitivities (solid histograms), scaled to three flavors, for a 14 day time window around the binary neutron star merger GW170817~\cite{ANTARES:2017bia} and the  all-flavor fluence predictions for millisecond magnetars from Fang \& Metzger \cite{Fang:2017tla}. 
Right: POEMMA cosmic neutrino sky coverage, without including the $\sim 50\%$ blocking effect of the Sun~\cite{Guepin2018}, at a  given day of the year for viewing angles to $\delta = 18.3^\circ$ below the limb. Figures from~\cite{Venters:2019xwi}. 
 \label{Neutr1}}
\end{figure}

Examples of neutrino rich astrophysical transient events include short and long gamma-ray bursts~\cite{Waxman:1997ti,Murase:2007yt,Kimura:2017kan}, gravitational wave events from neutron-star binary coalescence~\cite{Kimura:2017kan,Fang:2017tla}, black hole-black hole coalescence (BH-BH)~\cite{Kotera:2016dmp},  the birth of pulsars and magnetars~\cite{Fang:2014qva},  fast-luminous optical transients~\cite{Fang:2018hjp}, blazar flares~\cite{Rodrigues:2017fmu,IceCube:2018cha,IceCube:2018dnn}, tidal disruption events~\cite{Dai:2016gtz,Lunardini:2016xwi}, and possibly many other high-energy transients. 
Neutrinos, not gamma rays, may be the primary emission signal in some environments (e.g., cosmic ray acceleration in white dwarf-white dwarf (WD-WD) mergers~\cite{Xiao:2016man}). 
POEMMA can follow up these events and reach a neutrino fluence around $E_{\nu}^2J_{\nu} \ge  0.1 \ {\rm GeV}/ {\rm cm}^{-2}$ depending on the location of the sources (see Fig.~\ref{Neutr1} and~\cite{Venters:2019xwi}). 

\section{POEMMA Instrument and Mission}

The POEMMA concept (Fig.~\ref{fig1}) evolved from previous work on the OWL~\cite{Stecker:2004wt} and the JEM-EUSO~\cite{Adams:2013vea} designs, the CHANT concept~\cite{Neronov:2016zou}, and the sub-orbital payloads EUSO-SPB1~\cite{Wiencke:2017cfi} and EUSO-SPB2~\cite{Adams:2017fjh} (Extreme Universe Space Observatory on a Super Pressure Balloon, first and second flights). 

The POEMMA instrument design results from a work session at the Instrument Design Laboratory (IDL) of the Integrated Design Center (IDC) at the GSFC from July 31 to August 4, 2017. Following the IDL study, the POEMMA Study Team (ST) and key technical personnel at Marshall Space Flight Center (MSFC) de-scoped the design (simplifying and down-sizing the optics system and deployment mechanisms) to arrive at a balance between science objectives and resources.  The POEMMA mission design results from a work session at the Mission Design Laboratory (MDL) at the GSFC IDC from October 30 to November 3, 2017. (See \cite{NASApoemma} for  details of the POEMMA instrument and mission.)

POEMMA is composed of two identical space-based platforms that detect extreme energy particles by recording the signals generated by EASs in the dark side of the Earth's atmosphere.  The central element of each POEMMA observatory  is a high sensitivity low resolution photometer that measures and locates two types of emission from these EASs: the faint isotropic emission due to the fluorescence of atmospheric nitrogen excited by air shower particles, and the bright collimated Cherenkov emission from EASs directed at the POEMMA observatory.

\begin{figure}[tbp]
    \postscript{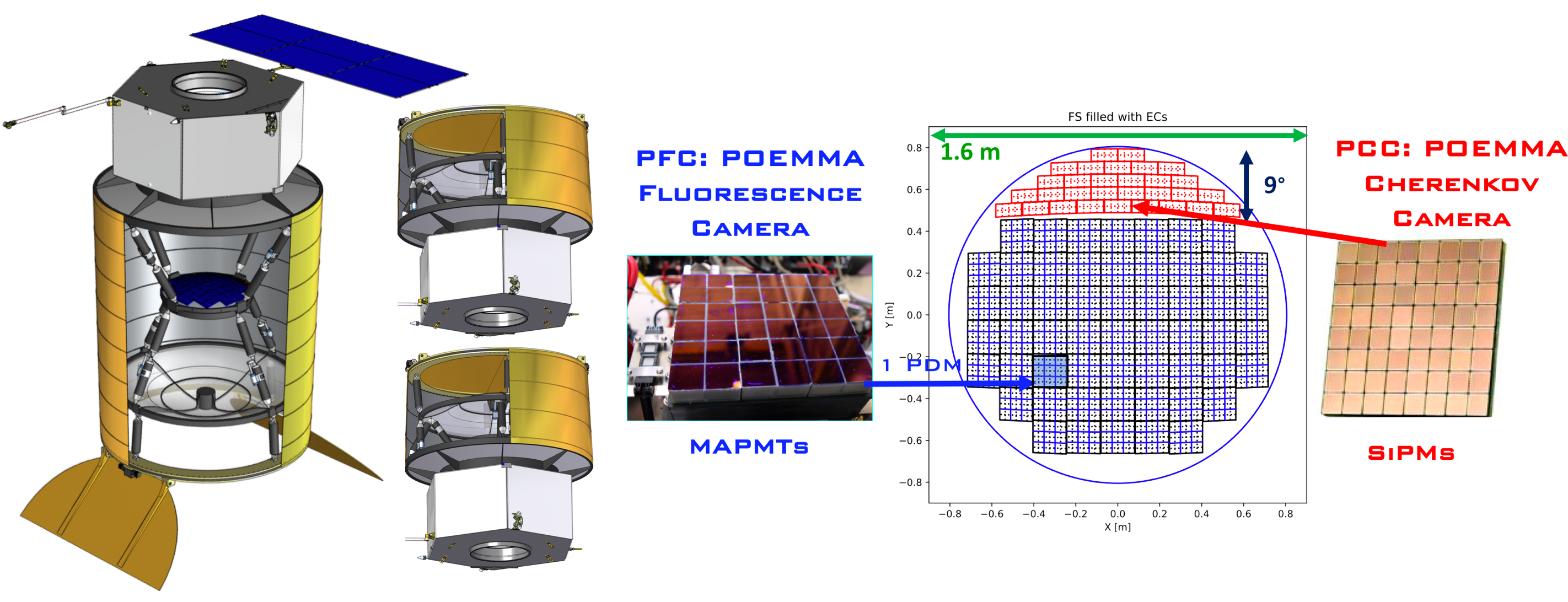}{1}
    \caption{Left: POEMMA instrument and spacecraft deployed and stowed for launch. Right: Layout of the photon sensors on the Focal Surface (PFC and PCC). \label{fig10}}
\end{figure}

POEMMA  photometers are designed for deployment after launch. A stowed configuration (see Fig.~\ref{fig10}) enables two identical satellites to be launched together on a single Atlas V rocket (see Fig.~\ref{fig1}). Space qualified mechanisms extend each instrument after launch to their deployed position to begin observations.
The instrument architecture incorporates a large number of identical parallel sensor chains that meet the high standards of a Class B mission. Aerospace grade components have been identified for key elements within these sensor chains to insure reliability for the mission. 

\noindent{\bf Optics:} The POEMMA photometer  is based on a Schmidt photometer with a large spherical primary mirror (4 m diameter), the aperture and a thin refractive aspheric aberration corrector lens (3.3 m diameter) at its center of curvature, and a convex spherical focal surface (1.61 m diameter). This particular system provides a large collection aperture (6.39 m$^2$) and a massive field-of-view (45$^\circ$ full FoV).   The diameter of the POEMMA primary mirror is set to fit the launch vehicle (Atlas V). 
The PSF of the POEMMA optics is much less than a pixel size. POEMMA's imaging requirement is  $10^4$ away from the diffraction limit, implying optical tolerances closer to a microwave dish than an astronomical telescope. 

\noindent{\bf Focal Surface:}
The focal surface (FS) of the POEMMA photometer consists of the POEMMA Fluorescence Camera (PFC), optimized for the fluorescence signals, and the POEMMA Cherenkov Camera (PCC), optimized for for Cherenkov signals (see Fig.~\ref{fig10}).
The PFC records the EAS videos in 1$\mu$s frames in the $300 \alt \lambda/{\rm nm} \alt 500$ wavelength band using multi-anode photomultiplier tubes (MAPMTs).  Each MAPMT has 64  (3x3 mm$^2$) pixels in an 8 x 8 array.  The PFC is composed of 1,980 MAPMTs containing a total of 126,720 pixels. The stereo videos of the EAS determines the energy, direction, and composition of the UHECR.
The PCC  uses silicon photomultipliers (SiPMs) and is optimized to observe in the  $300 \alt \lambda/{\rm nm} \alt 900$ wavelength band for Cherenkov emission of showers developing towards the observatory. The PCC covers 9$^\circ$ of the FoV from the edge  (see Fig.~\ref{fig10}). PCC sensors are solid-state SiPMs  assembled in arrays of 8 x 8 pixels with a total area of 31 x 31 mm$^2$. The PCC records EASs produced by UHECRs above the limb of the Earth and showers from $\tau$-lepton decays below the Earth's limb induced by $\nu_\tau$ interactions in the Earth. 

\noindent{\bf Data System:} The electronics and real-time software for the PFC and the PCC function independently so each can be optimized for their specific task. A single Data Processor (DP) will read out the data from both systems, store it locally and then transmit it to the spacecraft bus for transmission to the ground.

\noindent{\bf Mechanical Structure:} 
Structural analysis of the conceptual design shows a first fundamental frequency after deployment at 7.9 Hz, 60\% above the IDL goal of 5 Hz. An analysis of the stowed configuration during launch could not be completed due to the lack of design details of the launch vehicle. The deployment of the focal surface and corrector plate rely on one-time operations of actuators that drive the folded struts to reach full extension and then lock them into place for the duration of the mission. No further adjustment of the optical system is required due to the sizable tolerances afforded by the coarse EAS imaging requirements. 
 
\noindent{\bf Spacecraft bus:}
The spacecraft bus  is a hexagonal cylinder 1.55 m tall with an outside diameter of 3.37 m weighing 1,073 kg (including a 25\% contingency). Located behind the POEMMA mirror, it provides basic services to the observatory including on-orbit deployment, power, communications, attitude control, propulsion and avionics.  The avionics includes the command and data handling system (including the flight computer), the spacecraft clock that provides the precise timing we need for synchronization between the satellites, the gimbal drive electronics to steer the solar panels, and the control functions for all the deployment mechanisms to unfold the instrument once it reaches orbit. 

\noindent{\bf Mission Concept:}
After calibration, the instruments will be pointed close to the nadir in order to make stereo observations of the fluorescent light from cosmic ray extensive air showers (EASs) at the lowest energies. Once sufficient statistics have been acquired at the lowest energies, the satellites separation will be reduced to $\sim$ 30 km and the instruments will be pointed for limb observations via Cherenkov. Throughout the mission instruments will be re-oriented towards neutrino ToO directions following a transient event alert. During a ToO follow-up mode, measurements of fluorescence from EASs will continue utilizing the larger volumes of the atmosphere observed from the satellites to the limb. 

\noindent{\bf Launch Operations:}
The POEMMA satellites were designed for a dual-manifest on an Atlas V using the long payload fairing. The satellites will be launched into circular orbits at an inclination of 28.5 degrees and an altitude of 525 km, where they will remain until de-orbited at the end of the mission. 
The satellites are launched in a stowed configuration. Once on orbit, the corrector plate and focal surface will be deployed into their final position.  A solar array will be deployed from each spacecraft bus.

\noindent{\bf On-Orbit Operations:}
The satellites will orbit the Earth with a period of 95 minutes, or $\sim$15 times per day.  During observations the attitude of the satellites must be maintained within 0.1$^\circ$ with knowledge of the attitude to within 0.01$^\circ$. Events will be time-tagged with a relative accuracy within 25 ns between the satellites. 
The satellites have star trackers and sun sensors for accurate attitude knowledge.

\section{Technology Drivers and Partnerships}

The POEMMA concept relies on simple and proven technology. No new technologies are required to be developed for the mission. Technology maturations expected for the early part of the 2020s bring all components to high TRL. POEMMA team members are  working on the maturation of a few components through laboratory,  sub-orbital (EUSO-SPB1 and EUSO-SPB2), and spaceflight (mini-EUSO to be deployed in the ISS in 2019/20) testing. In particular, EUSO-SPB2 is being built with the same focal surface components of POEMMA and plans to fly on a NASA super-pressure balloon in 2022.

\noindent{\bf Focal surface: }  The IDL concluded that the PFC could be built today with existing technology. The PCC SiPM-based section would benefit from space qualification of SiPM arrays and silicon photo-diodes, packaging studies for arrays and diodes to conform to the focal surface, and space qualification of the SiPM ASIC.

\noindent{\bf Mechanical:} The shutter doors must operate during the entire mission and are the highest mission risk. This risk can be reduced through iterative design and testing cycles. Alternative options to the door system such as active pointing away from bright sources should be studied  in the preliminary design phase. 

\noindent{\bf Optics:}  POEMMA optics meets science requirements with no issue for manufacturing. Studies to space qualify UV coatings can increase the performance. 
\medskip

\noindent{\bf Organization, Partnerships, and Current Status}
The POEMMA Collaboration includes members of the JEM-EUSO, OWL, Auger, TA, CTA, and Fermi collaborations. The JEM-EUSO collaboration (https://jemeuso.riken.jp) includes members from 84 institutions from 16 countries committed to an international partnership to build POEMMA.

\begin{figure}[tbp]
    \postscript{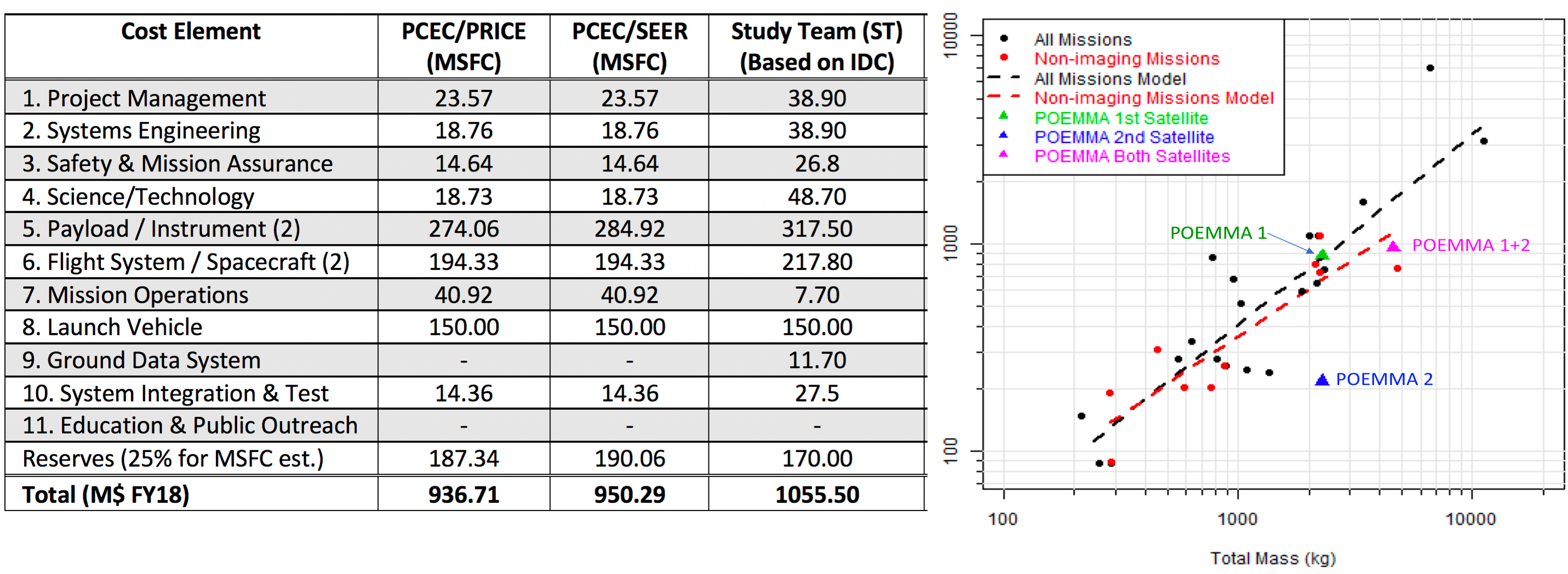}{0.99}
    \caption{Left: Cost estimates  in millions of FY18 dollars.
Right: POEMMA costs compared to past missions. One unit of POEMMA aligns well with past costs while the second unit adds 25\% to the total cost.
(Historical Missions include: ACTS, CALIPSO, Cloudsat, FUSE, GALEX, Herschel, HST, ICESat, IRAS, IUE, JWST, Kepler, MRO/HiRISE, OAO-2/CEP, OAO-3/PEP, OAO-B/GEP, Planck, SOHO/EIT, Spitzer, SECCHI A, SWAS, TDRS-1, TDRS-7, TRACE, WIRE, WISE, WMAP.) \label{CostComp}}
\end{figure}

\section{Cost and Schedule}

We estimate POEMMA's total life cycle cost between \$0.94 and \$1.05 billion in FY18 dollars, including reserves and the launch vehicle according to the ground rules of the NASA probe study. The table in Fig. \ref{CostComp} shows the high level overview of three bounding cost estimates. The POEMMA Study Team (ST) estimate is based on the GSFC IDC parametric cost model of the first iteration of the instrument concept. The initial concept was costed to be well over the \$1B Probe-class limit and was de-scoped to the concept presented here.  In addition, two more parametric cost models were constructed by the Engineering Cost Office (ECO) at MSFC.  Assumptions common to these estimates were for a Class B mission, 4-6 year development schedule, launch in the 2027-2029 time frame and \$150M for launch services. Reserves were set by NASA Headquarters (HQ) to be \$170M.  (MSFC ECO typically sets reserves to 25\% at minimum.) The MSFC ECO cost models were constructed (in January-February 2018) at the subsystem level utilizing the Master Equipment List (MEL) from the  IDL for the payload/instrument adjusted by the ST as required by the concept de-scope, and the MEL from the  MDL for the spacecraft. The three total cost estimates have not explored sensitivities to mass and schedule. 

The ECO team compared their final value to total mass versus overall cost data for past missions as shown in Fig. \ref{CostComp} right. The cost for a single POEMMA spacecraft (green) falls on trend line model for all missions (dashed black). 
The low resolution requirements for the POEMMA optics is comparable to non-imaging missions (red dots and red-dashed line).
Given the front-loaded costs for a Class B mission, the second POEMMA satellite costs significantly less than the first. The GSFC IDC and MSFC ECO results confirm that the spacecraft exact copy costs 25\% of the first POEMMA satellite for Class A/B missions (blue triangle). 
This is a factor of two less than is customary for Class C/D missions since payload management, systems engineering, and safety/mission assurance  costs are much higher for Class A/B missions and are not recurring for exact spacecraft copies.
The complete POEMMA mission (magenta triangle) falls just below the non-imaging mission cost-vs-mass trend line model (red dashed). 

\begin{figure}[tbp]
    \postscript{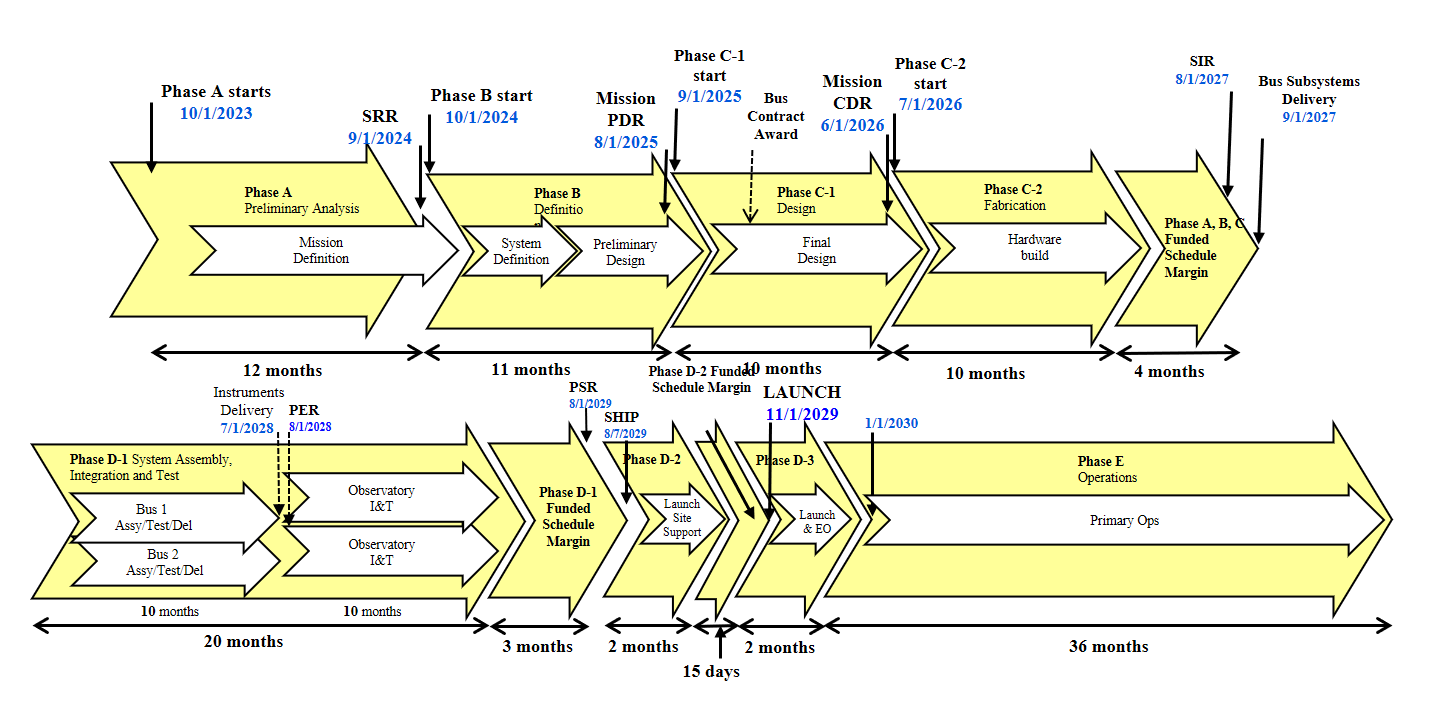}{0.9}
    \caption{Mission Schedule \label{Sched}}
\end{figure}

The timeline for the mission was developed at the MDL as shown in Fig.~\ref{Sched}. The launch date was selected by NASA Headquarters for 2029 which led to a phase A start at 2023. The MDL accessed the schedule to be aggressive but feasible, given the modularity and duplicate systems of both instruments and spacecraft.

\section{Summary}
POEMMA is a unique probe-class mission designed to answer fundamental open questions in the multi-messenger domain starting the next decade. Community driven white-papers on the UHECR science \cite{Sarazin:2019} and astrophysical neutrino science \cite{Ackermann:2019a,Ackermann:2019b} describe the scientific challenges that POEMMA will address. Other relevant science white papers submitted to Astro2020 include (NAS response ID): 107, 147, 206, 239, 253, 263, 265, 272, 275, 277, 377, 439.
The multi-messenger domain began over the last decade and it will flourish during the 2020s and 2030s with the capability provided by POEMMA.


\begin{thebibliography}{99}

\bibitem{NASApoemma} NASA 020 Decadal Survey Planning site: https://science.nasa.gov/astrophysics/2020-decadal-survey-planning includes the POEMMA design report funded by NASA Grant Number: NNX17AJ82G 

https://smd-prod.s3.amazonaws.com/science-pink/s3fs-public/atoms/files/
1\_POEMMA\_Study\_Rpt\_0.pdf

\bibitem{UHECRpoemma}  
L. A. Anchordoqui {\it et al.},
{\color{rossoCP3} UHECRs with POEMMA} 
arXiv:1907.03694 
 
\bibitem{Ackermann:2012vca}
M. Ackermann, {\it et al.} 
   {\color{rossoCP3} GeV Observations of Star-forming Galaxies with  Fermi-LAT}
  Astrophys. J., 755 (2012) 164, arXiv:1206.1346

\bibitem{Aab:2018chp} 
  A.~Aab {\it et al.} [Pierre Auger Collaboration],
   {\color{rossoCP3} An Indication of anisotropy in arrival directions of ultra-high-energy cosmic rays through comparison to the flux pattern of extragalactic gamma-ray sources},
  Astrophys.\ J.\  {\bf 853}, no. 2, L29 (2018)
  doi:10.3847/2041-8213/aaa66d
  [arXiv:1801.06160 [astro-ph.HE]].

\bibitem{Abraham:2010zz}
  J.~Abraham {\it et al.}  [Pierre Auger  Collaboration],
  {\color{rossoCP3} Trigger and aperture of the surface detector array of the Pierre Auger
  Observatory},
  Nucl.\ Instrum.\ Meth.\  A {\bf 613}, 29 (2010).

\bibitem{Abraham:2009pm}
  J.~A.~Abraham {\it et al.}  [Pierre Auger Collaboration],
  {\color{rossoCP3} The fluorescence detector of the Pierre Auger Observatory},
  Nucl.\ Instrum.\ Meth.\  A {\bf 620}, 227 (2010)
  [arXiv:0907.4282].

\bibitem{Aab:2017njo} 
  A.~Aab {\it et al.} [Pierre Auger Collaboration],
   {\color{rossoCP3} The Pierre Auger Observatory: Contributions to the 35th International Cosmic Ray Conference (ICRC 2017)},
  arXiv:1708.06592 [astro-ph.HE].

\bibitem{AbuZayyad:2012kk} 
  T.~Abu-Zayyad {\it et al.} [Telescope Array Collaboration],
   {\color{rossoCP3} The surface detector array of the Telescope Array experiment},
  Nucl.\ Instrum.\ Meth.\ A {\bf 689}, 87 (2013)
  doi:10.1016/j.nima.2012.05.079
  [arXiv:1201.4964 [astro-ph.IM]].
  
    
\bibitem{Tokuno:2012mi} 
  H.~Tokuno {\it et al.},
   {\color{rossoCP3} New air fluorescence detectors employed in the Telescope Array experiment},
  Nucl.\ Instrum.\ Meth.\ A {\bf 676}, 54 (2012)
  doi:10.1016/j.nima.2012.02.044
  [arXiv:1201.0002 [astro-ph.IM]].

\bibitem{Kotera:2011cp}
  K.~Kotera and A.~V.~Olinto,
    {\color{rossoCP3}  The astrophysics of ultrahigh energy cosmic rays},
  Ann.\ Rev.\ Astron.\ Astrophys.\  {\bf 49}, 119 (2011)
  doi:10.1146/annurev-astro-081710-102620
  [arXiv:1101.4256 [astro-ph.HE]].

\bibitem{Anchordoqui:2018qom}
  L.~A.~Anchordoqui,
  {\color{rossoCP3}   Ultra-high-energy cosmic rays},
  arXiv:1807.09645 [astro-ph.HE].

\bibitem{Batista:2019}
R. A. Batista et al., 
{\color{rossoCP3} Open Questions in Cosmic-Ray Research at Ultrahigh Energies}
 Frontiers in Astronomy and Space Sciences 2019.
 
 
 \bibitem{Blasi:2000xm} 
  P.~Blasi, R.~I.~Epstein and A.~V.~Olinto,
    {\color{rossoCP3} Ultrahigh-energy cosmic rays from young neutron star winds},
  Astrophys.\ J.\  {\bf 533}, L123 (2000)
  doi:10.1086/312626
  [astro-ph/9912240].

\bibitem{Fang:2012rx} 
  K.~Fang, K.~Kotera and A.~V.~Olinto,
    {\color{rossoCP3} Newly-born pulsars as sources of ultrahigh energy cosmic rays},
  Astrophys.\ J.\  {\bf 750}, 118 (2012)
  doi:10.1088/0004-637X/750/2/118
  [arXiv:1201.5197 [astro-ph.HE]].

\bibitem{Fang:2013cba} 
  K.~Fang, K.~Kotera and A.~V.~Olinto,
    {\color{rossoCP3} Ultrahigh energy cosmic ray nuclei from extragalactic pulsars and the effect of their Galactic counterparts},
  JCAP {\bf 1303}, 010 (2013)
  doi:10.1088/1475-7516/2013/03/010
  [arXiv:1302.4482 [astro-ph.HE]].

\bibitem{Biermann:1987ep} 
  P.~L.~Biermann and P.~A.~Strittmatter,
   {\color{rossoCP3}  Synchrotron emission from shock waves in active galactic nuclei},
  Astrophys.\ J.\  {\bf 322}, 643 (1987).
  doi:10.1086/165759

\bibitem{Rachen:1992pg} 
  J.~P.~Rachen and P.~L.~Biermann,
   {\color{rossoCP3}  Extragalactic ultrahigh-energy cosmic rays 1: Contribution from hot spots in FR-II radio galaxies},
  Astron.\ Astrophys.\  {\bf 272}, 161 (1993)
  [astro-ph/9301010].

\bibitem{Eichmann:2017iyr} 
  B.~Eichmann, J.~P.~Rachen, L.~Merten, A.~van Vliet and J.~Becker Tjus,
 {\color{rossoCP3}  Ultra-high-energy cosmic rays from radio galaxies},
  JCAP {\bf 1802}, no. 02, 036 (2018)
  doi:10.1088/1475-7516/2018/02/036
  [arXiv:1701.06792 [astro-ph.HE]].


\bibitem{Anchordoqui:1999cu} 
  L.~A.~Anchordoqui, G.~E.~Romero and J.~A.~Combi,
 {\color{rossoCP3}   Heavy nuclei at the end of the cosmic ray spectrum?},
  Phys.\ Rev.\ D {\bf 60}, 103001 (1999)
  doi:10.1103/PhysRevD.60.103001
  [astro-ph/9903145].

\bibitem{Anchordoqui:2018vji} 
  L.~A.~Anchordoqui,
   {\color{rossoCP3}  Acceleration of ultrahigh-energy cosmic rays in starburst superwinds},
  Phys.\ Rev.\ D {\bf 97}, no. 6, 063010 (2018)
  doi:10.1103/PhysRevD.97.063010
  [arXiv:1801.07170 [astro-ph.HE]].

\bibitem{Waxman:1995vg} 
  E.~Waxman,
    {\color{rossoCP3} Cosmological gamma-ray bursts and the highest energy cosmic rays},
  Phys.\ Rev.\ Lett.\  {\bf 75}, 386 (1995)
  doi:10.1103/PhysRevLett.75.386
  [astro-ph/9505082].

\bibitem{Vietri:1995hs} 
  M.~Vietri,
    {\color{rossoCP3} On the acceleration of ultrahigh-energy cosmic rays in gamma-ray bursts},
  Astrophys.\ J.\  {\bf 453}, 883 (1995)
  doi:10.1086/176448
  [astro-ph/9506081].
  
\bibitem{Muzio:2019leu} 
  M.~S.~Muzio, M.~Unger and G.~R.~Farrar,
   {\color{rossoCP3} Progress towards characterizing ultrahigh energy cosmic ray sources,}
  arXiv:1906.06233 
 
\bibitem{Guepin2018}
 C.~Gu$\acute{\rm e}$pin, F.~Sarazin, J.~Krizmanic, J.~Loerincs, A.~Olinto, and A.~Piccone,
 {\color{rossoCP3}   Geometrical Constraints of Observing Very High Energy Earth-Skimming Neutrinos from Space},
 [arXiv:1812.07596 [astro-ph.IM]]

\bibitem{Reno2019a}
M.~H.~Reno, J.~F.~Krizmanic and T.~M.~Venters,
  {\color{rossoCP3} Cosmic tau neutrino detection via Cherenkov signals from air showers from Earth-emerging taus}, [arXiv:1902.11287 [astro-ph.HE]].

\bibitem{Venters:2019xwi}
T. M. Venters, M. H. Reno, J. F. Krizmanic, L. A.  Anchordoqui, C.  GuŽpin, and A. V. Olinto, 
{\color{rossoCP3}  POEMMA's Target of Opportunity Sensitivity to Cosmic Neutrino Transient Sources}, (2019)
arXiv:1906.07209

\bibitem{Meszaros:2017fcs}
P. Meszaros, 
    {\color{rossoCP3} Astrophysical Sources of High Energy Neutrinos in the IceCube Era},
Ann. Rev. Nucl. Part. Sci. 67, (2017), 45-67; arXiv:1708.03577
     
\bibitem{Ackermann:2019ows}
M. Ackermann, et al., 
    {\color{rossoCP3} Astrophysics Uniquely Enabled by Observations of High-Energy Cosmic Neutrinos}",
(2019), arXiv:1903.04334",
   
\bibitem{Waxman:1997ti}
E. Waxman and J. N. Bahcall
    {\color{rossoCP3} High-energy neutrinos from cosmological gamma-ray burst fireballs},
Phys. Rev. Lett. 78 (1997) 2292-2295", astro-ph/9701231

\bibitem{Murase:2007yt}
K. Murase,
    {\color{rossoCP3} High energy neutrino early afterglows gamma-ray bursts revisited},
Phys. Rev. D76 (2007) 123001, arXiv:0707.1140


\bibitem{Kimura:2017kan} 
  S.~S.~Kimura, K.~Murase, P.~MŽsz‡ros and K.~Kiuchi,
  {\color{rossoCP3}High-Energy Neutrino Emission from Short Gamma-Ray Bursts: Prospects for Coincident Detection with Gravitational Waves,}
  Astrophys.\ J.\  {\bf 848}, no. 1, L4 (2017)
  doi:10.3847/2041-8213/aa8d14
  [arXiv:1708.07075 [astro-ph.HE]].

\bibitem{Fang:2017tla} 
  K.~Fang and B.~D.~Metzger,
  {\color{rossoCP3}High-Energy Neutrinos from Millisecond Magnetars formed from the Merger of Binary Neutron Stars,}
  Astrophys.\ J.\  {\bf 849}, no. 2, 153 (2017)
  [Astrophys.\ J.\  {\bf 849}, 153 (2017)]
  doi:10.3847/1538-4357/aa8b6a
  [arXiv:1707.04263 [astro-ph.HE]].
 
 
\bibitem{Kotera:2016dmp}
K. Kotera and J. Silk
    {\color{rossoCP3} Ultrahigh Energy Cosmic Rays and Black Hole Mergers}
Astrophys. J. 823 (2016) 2, L29, arXiv:1602.06961

\bibitem{Fang:2014qva}
K. Fang, 
    {\color{rossoCP3} High-Energy Neutrino Signatures of Newborn Pulsars In the Local Universe},
JCAP 1506 (2015) 06, arXiv:1411.2174

\bibitem{Fang:2018hjp} 
  K.~Fang, B.~D.~Metzger, K.~Murase, I.~Bartos and K.~Kotera,
  {\color{rossoCP3} Multimessenger Implications of AT2018cow: High-Energy Cosmic Ray and Neutrino Emissions from Magnetar-Powered Super-Luminous Transients,}
  arXiv:1812.11673 [astro-ph.HE].


\bibitem{Rodrigues:2017fmu}
X. Rodrigues,A. Fedynitch, S. Gao, D.  Boncioli,  and W. Winter,
    {\color{rossoCP3} Neutrinos and Ultra-High-Energy Cosmic-Ray Nuclei from Blazars}, Astrophys. J. 854 (2018) 1, 54,
arXiv:1711.02091

\bibitem{IceCube:2018cha} 
  M.~G.~Aartsen {\it et al.} [IceCube Collaboration],
  {\color{rossoCP3}  Neutrino emission from the direction of the blazar TXS 0506+056 prior to the IceCube-170922A alert},
  Science {\bf 361}, no. 6398, 147 (2018).
  doi:10.1126/science.aat2890
 [arXiv:1807.08794 [astro-ph.HE]].


\bibitem{IceCube:2018dnn}
M. G. Aartsen, M. G. et al.,
    {\color{rossoCP3} Multimessenger observations of a flaring blazar coincident with high-energy neutrino IceCube-170922A},
Science 361 (2018) 6398, eaat1378, arXiv:1807.08816

\bibitem{Dai:2016gtz}
L. Dai and K. Fang,
    {\color{rossoCP3} Can tidal disruption events produce the IceCube neutrinos?},
Mon. Not. Roy. Astron. Soc., 469 (2017) 2, 1354-1359, arXiv:1612.00011

\bibitem{Lunardini:2016xwi}
C. Lunardini and W. Winter,
    {\color{rossoCP3}High Energy Neutrinos from the Tidal Disruption of Stars},
Phys. Rev.,
D95,
2017,
12,
123001,
arXiv:1612.03160


\bibitem{Xiao:2016man}
D. Xiao, P. M\'esz\'aros, K. Murase, and Z.-g. Dai,
    {\color{rossoCP3} High-Energy Neutrino Emission from White Dwarf Mergers},
Astrophys. J.
832,
(2016)
20,
arXiv:1608.08150

\bibitem{Stecker:2004wt} 
  F.~W.~Stecker, J.~F.~Krizmanic, L.~M.~Barbier, E.~Loh, J.~W.~Mitchell, P.~Sokolsky and R.~E.~Streitmatter,
    {\color{rossoCP3} Observing the ultrahigh-energy universe with OWL eyes},
  Nucl.\ Phys.\ Proc.\ Suppl.\  {\bf 136C}, 433 (2004)
  doi:10.1016/j.nuclphysbps.2004.10.027
  [astro-ph/0408162].

\bibitem{Adams:2013vea} 
  J.~H.~Adams {\it et al.} [JEM-EUSO Collaboration],
    {\color{rossoCP3} An evaluation of the exposure in nadir observation of the JEM-EUSO mission},
  Astropart.\ Phys.\  {\bf 44}, 76 (2013)
  doi:10.1016/j.astropartphys.2013.01.008
  [arXiv:1305.2478 [astro-ph.HE]].

\bibitem{Neronov:2016zou} 
  A.~Neronov, D.~V.~Semikoz, L.~A.~Anchordoqui, J.~Adams and A.~V.~Olinto,
    {\color{rossoCP3} Sensitivity of a proposed space-based Cherenkov astrophysical-neutrino telescope},
  Phys.\ Rev.\ D {\bf 95}, no. 2, 023004 (2017)
  doi:10.1103/PhysRevD.95.023004
  [arXiv:1606.03629 [astro-ph.IM]].

\bibitem{Wiencke:2017cfi} 
  L.~Wiencke {\it et al.} [JEM-EUSO Collaboration],
    {\color{rossoCP3} EUSO-SPB1 mission and science},
  PoS ICRC {\bf 2017}, 1097 (2018).
  doi:10.22323/1.301.1097

\bibitem{Adams:2017fjh} 
  J.~H.~Adams {\it et al.},
    {\color{rossoCP3} White paper on EUSO-SPB2},
  arXiv:1703.04513 [astro-ph.HE].

\bibitem{ANTARES:2017bia} 
  A.~Albert {\it et al.} [ANTARES and IceCube and Pierre Auger and LIGO Scientific and Virgo Collaborations],
  {\color{rossoCP3}Search for High-energy Neutrinos from Binary Neutron Star Merger GW170817 with ANTARES, IceCube, and the Pierre Auger Observatory,}
  Astrophys.\ J.\  {\bf 850}, no. 2, L35 (2017)
  doi:10.3847/2041-8213/aa9aed
  [arXiv:1710.05839 [astro-ph.HE]].

\bibitem{Sarazin:2019}
F. Sarazin et al., 
 {\color{rossoCP3} What is the Nature and Origin of the Highest-Energy Particles in the Universe?},
 white paper for the 2020 Decadal Survey, 2019

\bibitem{Ackermann:2019a}
M.~Ackermann et al., 
{\color{rossoCP3} Astrophysics Uniquely Enabled by Observations of High-Energy Cosmic Neutrinos},
Astro2020 Science White Paper (2019)

\bibitem{Ackermann:2019b}
M.~Ackermann et al., 
{\color{rossoCP3} Fundamental Physics with High-Energy Cosmic Neutrinos}
Astro2020 Science White Paper (2019).




\end{thebibliography}
\end{document}